\documentclass[12pt,english]{iopart}
\usepackage[T1]{fontenc}
\usepackage[utf8x]{inputenc}
\usepackage{cite}
\usepackage{graphicx}
\usepackage{amssymb}

 \usepackage{amsfonts}
 \usepackage{epsfig}
 \usepackage{subfigure}
 \usepackage{graphicx}
 \usepackage{float}

\makeatletter
\@ifundefined{textcolor}{}
{
 \definecolor{BLACK}{gray}{0}
 \definecolor{WHITE}{gray}{1}
 \definecolor{RED}{rgb}{1,0,0}
 \definecolor{GREEN}{rgb}{0,1,0}
 \definecolor{BLUE}{rgb}{0,0,1}
 \definecolor{CYAN}{cmyk}{1,0,0,0}
 \definecolor{MAGENTA}{cmyk}{0,1,0,0}
 \definecolor{YELLOW}{cmyk}{0,0,1,0}
 }

\makeatother

\usepackage{babel}

\begin{document}
\title[Binary bosonic mixture in a state-dependent honeycomb lattice]
{Beyond mean-field study of a binary bosonic mixture in a state-dependent honeycomb lattice}

\author{Lushuai Cao $^{1,3}$, Sven Kr\"onke$^{1}$, Jan Stockhofe$^{1}$, Peter Schmelcher$^{1,3}$,
Juliette Simonet$^{2}$, Klaus Sengstock $^{1,2}$, Dirk-S\"oren L\"{u}hmann$^{2}$}
\address{$^1$ Zentrum f\"{u}r Optische Quantentechnologien, Universit\"{a}t Hamburg, Luruper Chaussee 149, D-22761 Hamburg, Germany}
\address{$^2$ Institut f\"{u}r Laserphysik, Universit\"{a}t Hamburg, Luruper Chaussee 149, D-22761 Hamburg, Germany}
\address{$^3$ The Hamburg Centre for Ultrafast Imaging, Luruper Chaussee 149, D-22761 Hamburg, Germany}
\ead{
\mailto{lcao@physnet.uni-hamburg.de},
\mailto{sven.kroenke@physnet.uni-hamburg.de}, 
\mailto{jstockho@physnet.uni-hamburg.de}, 
\mailto{pschmelc@physnet.uni-hamburg.de},
\mailto{jsimonet@physik.uni-hamburg.de},
\mailto{ksengsto@physnet.uni-hamburg.de},
\mailto{dluehman@physnet.uni-hamburg.de}}

\begin{abstract}
We investigate a binary mixture of bosonic atoms loaded into a state-dependent honeycomb lattice. For this system, the emergence of a so-called twisted-superfluid ground state was experimentally observed in 
[Soltan-Panahi \textit{et al.}, Nat. Phys. {\bf 8}, 71 (2012)]. Theoretically, the origin of this effect is not understood.
We perform numerical simulations of an extended Bose-Hubbard model adapted to the experimental parameters employing
the Multi-Layer Multi-Configuration Time-Dependent Hartree method for Bosons.
Our results confirm the overall applicability of mean-field theory within the relevant parameter range.
Beyond this, we provide a detailed analysis of correlation effects correcting the mean-field result.
These have the potential to induce asymmetries in single shot time-of-flight measurements, but we find no indication of the patterns characteristic of the twisted superfluid.
We comment on the restrictions of our model and possible extensions.
\end{abstract}


\maketitle
\section{Introduction}
 The experimental realization of Bose-Einstein condensates (BECs) in dilute
gases  of alkali atoms in 1995 \cite{BEC1,BEC2,BEC3} has triggered extensive studies on ultracold atomic systems. 
Present day experimental techniques allow for highly precise control of ultracold atomic ensembles, 
including an almost perfect decoupling from the environment as well as a free design of the confining potential landscape
\cite{lattice1,lattice2} and even tunability of the inter-atom interactions, via Feshbach  \cite{Feshbach,Feshbach-magnetical,Feshbach-optical} and confinement
induced resonances \cite{CIR,CIR2,CIR3,CIR4,CIR5}. 
This exceptional degree of controllability suggests to use ultracold atom systems for the purpose of quantum simulation.
Phenomena to be potentially simulated with ultracold atoms in the laboratory come from fields as diverse as cosmology \cite{KZM1,KZM2,KZM3,Sakharov} or strongly correlated condensed matter systems \cite{Mott1,Mott2,anderson1,anderson2,Hall1}, to name only a few. 
Importantly, internal (such as spin) degrees of freedom of the system to be simulated can be mimicked using different species of atoms.
Such multi-species ultracold ensembles, including Bose-Bose \cite{bbmixture1,bbmixture2}, Fermi-Fermi \cite{ffmixture1,ffmixture2} and even Bose-Fermi mixtures \cite{bfmixture1,bfmixture2,bfmixture3,bfmixture4,bfmixture5}, have been extensively studied.
  
Among the various observations made in ultracold atomic mixtures, an intriguing observation of the so-called twisted superfluid (TSF) state is reported for a two-component bosonic mixture loaded into a state-dependent honeycomb lattice (SDHL) \cite{TSF_exp}. Experimentally, it is found that for a range of parameters of the SDHL potential the ground state of the binary atomic system has unusual properties, setting it apart from an ordinary superfluid phase. In such a lattice geometry, for a single spin state, the free-space momentum distribution displays a six-fold rotational symmetry in the first order Bragg peaks observed after time-of-flight. For a mixture of two spin states, two features characteristic for the TSF phase are observed in the quasi-momentum distribution:
   \begin{itemize}
   \item[(P1)] A reduced three-fold symmetry is observed for each spin state, appearing as an alternating pattern in the first-order momentum peaks. The six first order Bragg peaks can thus be grouped in two sets, each forming an equilateral triangle.
   \item[(P2)] It is furthermore observed that the two spin state show a complementary momentum distribution, populating different triangles. The six-fold symmetry expected for the lattice geometry is then restored in the summed momentum distribution of the two spin states.
  \end{itemize}

Theoretically, the mechanism driving the emergence of such a TSF state is unclear.
A recent mean-field study \cite{TSF_mf} did not find signatures of the type (P1) or (P2),
speculating at the same time about the relevance of beyond mean-field effects.
In this work, we aim to clarify this point by performing full many-body quantum simulations of an extended Bose-Hubbard model beyond the mean-field
approximation to investigate the ground state of binary bosonic mixtures in the SDHL, looking in particular for the above TSF signatures.
Our investigation shows that in the small lattice depth regime, where the emergence of TSF was found in the experiment \cite{TSF_exp}, 
the system is indeed well described by mean-field theory.
Correcting the mean-field approximation, we observe inter-species quantum correlations with a weak but finite amplitude.
We demonstrate that these correlations can modify the population distribution of the first order Bragg peaks in free-momentum space, 
however our findings do not confirm the presence of properties (P1) and (P2).
These results indicate that within the lowest band extended Bose-Hubbard model the twisted superfluid effect cannot be ascribed to quantum correlation effects beyond mean-field.
The main limitations of our analysis are the restriction to the lowest Bloch band and the finite domain size of the numerical simulations. 
Also higher order tunneling processes such as the density-induced tunneling \cite{Wannier} 
that are not contained in our Bose-Hubbard Hamiltonian might play a role.

The paper is organized as follows: in chapter 2 we briefly describe the experimental setup and
the Hamiltonian corresponding to our theoretical model.
The numerical method Multi-Layer Multi-Configuration Time-Dependent
Hartree method for Bosons (ML-MCTDHB) is introduced in chapter 3. 
In chapter 4 we present the results of the simulations, and a summary as well as a discussion is given in chapter 5.

\section{Setup}\label{setup}
In the experiment \cite{TSF_exp}, the ground state properties of binary bosonic
mixtures loaded into a two-dimensional SDHL are investigated. The atomic samples consist of
${}^{87}{\rm Rb}$ atoms in different spin states $|F,m_F\rangle$ from the
hyperfine manifolds $F=1,2$ with Zeeman sublevels $m_F$. The atoms experience a
state-dependent potential generated by the intersection of three coplanar
linearly polarized laser beams at an angle of $120^\circ$, with the
polarization of each laser beam lying in the intersection plane. 
The quantization axis of the system is defined by a homogeneous magnetic field.
Orientating the
magnetic field perpendicular to the lattice plane results in
alternating circular polarization on lattice sites and thereby gives
the following SDHL
potential \cite{Wannier}:
\begin{equation}\label{SDHL}
 V({\bf x})=-V_0[6-2\sum_{i=1}^3\cos({\bf b}_i{\bf x})+\sqrt{3}(-1)^F
m_F \eta\sum_{i=1}^3 \sin({\bf b}_i{\bf x})],
\end{equation}
where $V_0$ and $\lambda$ denote the lattice depth and laser wavelength,
respectively. The vectors ${\bf b}_1=\frac{2\pi}{\lambda}(\sqrt{3},0)$ and ${\bf
b}_2=\frac{2\pi}{\lambda}(-\frac{\sqrt{3}}{2},\frac{3}{2})$ span the reciprocal
Bravais lattice while ${\bf b}_3=-{\bf b}_1-{\bf b}_2$.
The weight of the state-dependent term is given by the dimensionless
constant $\eta=0.13$ for the experimental values \cite{Wannier}. In the third
spatial dimension, the atomic motion is frozen by a deep one-dimensional
lattice, creating effectively a stack of planar samples taken to be fully
decoupled from each other.

\begin{figure}
 \centering
\subfigure[]{

\includegraphics[width=0.4\textwidth]
{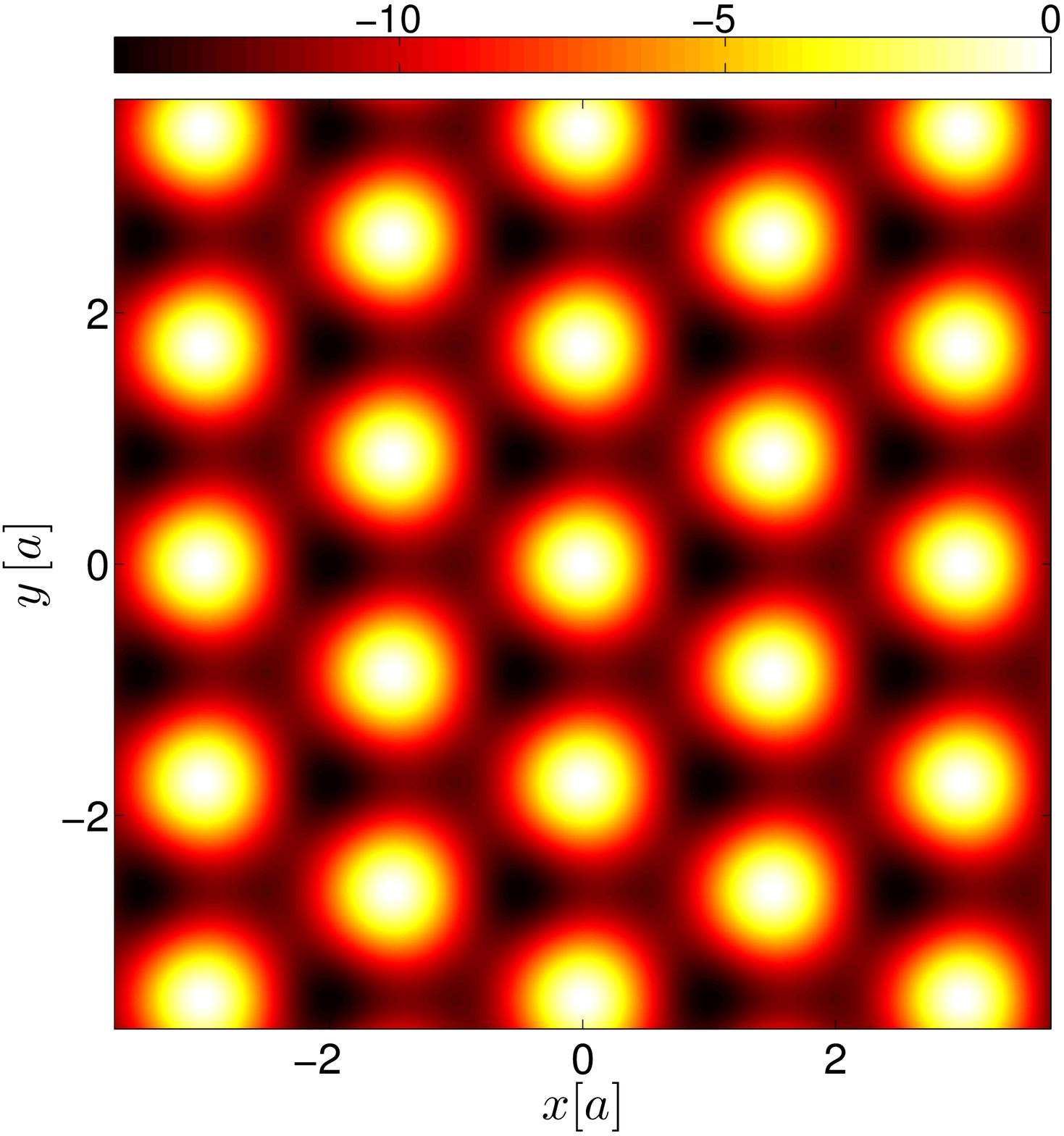}\label{figure11}
}
\subfigure[]{

 \includegraphics[width=0.4\textwidth]
 {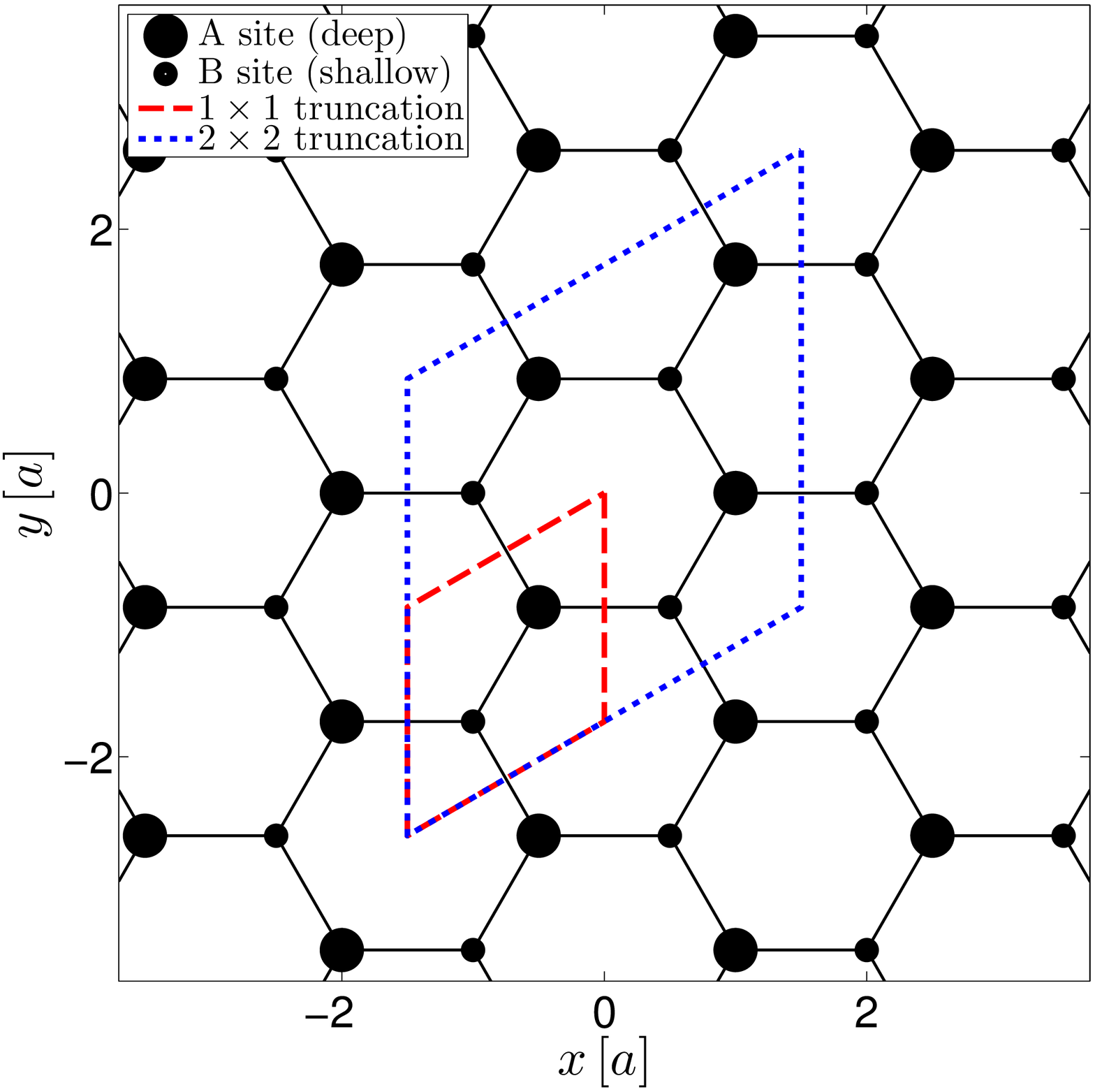}\label{figure12}
}
\caption{(a) The SDHL potential landscape $V(x,y)$ and (b) a schematic sketch of the
lattice sites for $\uparrow$ bosons. The potential in (a) is calculated
from equation. (\ref{SDHL}) at a lattice depth $V_0=1.5\,E_R$. In figure (b),
the big/small spots denote the A/B lattice sites, respectively, which are deep/shallow sites for $\uparrow$ bosons. 
The boxes with red and blue dashed lines indicate the spatial regions covered by our $1\times1$ and $2\times2$ unit cell truncations.}
\label{figure1}
\end{figure}

In the following we will focus on a bosonic mixture of the two spin states
$|1,1\rangle$ and $|1,-1\rangle$, denoted as $\uparrow$ and
$\downarrow$ spins, for which the emergence of the TSF phase has been observed
experimentally \cite{TSF_exp}. As shown for the $\uparrow$ species in figure
\ref{figure11}, the SDHL features a non-trivial unit cell of two, in general,
inequivalent sites referred to as A and B: While the A/B sites correspond to
deep/shallow sites for the $\uparrow$ species, they exactly interchange their
roles for the $\downarrow$ species. We introduce $a=\frac{2}{3 \sqrt{3}}\lambda$
denoting the distance between a neighboring $A-B$ pair.

\begin{figure}
\centering
 \includegraphics[width=0.4\textwidth]
{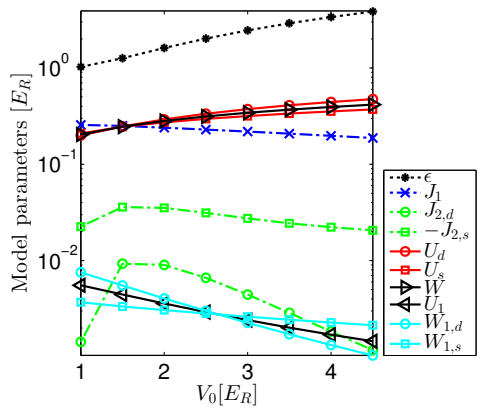}
\caption{Parameters of the extended BH model as a function of
the lattice depth, calculated according to \cite{Wannier} for the setup of the experiment \cite{TSF_exp}. 
}
\label{figure2}
\end{figure}

In our work, the numerical simulations are based on an extended Bose-Hubbard
(BH) model for the binary bosonic mixture in the SDHL. The construction of this
extended Bose-Hubbard model is optimized in the sense that the amplitudes of
leading-order neglected processes (for instance the density-induced tunneling)
are minimized by the choice of optimal Wannier functions \cite{Wannier}. This
optimization process gives rise to a single-band\footnote{Our
approximation can equivalently be viewed as an extended two-band BH model if one
regards the unit cells (A/B double wells) as elementary sites.}
BH model, in which
only one Wannier function is assigned to each site, while higher-band states are
solely used for constructing these optimized Wannier functions \cite{Wannier}.
The
resulting extended BH Hamiltonian contains not only the conventional terms, i.e.
site offset energies, nearest-neighbor (NN) hopping and on-site interactions,
but also next-to-nearest-neighbor (NNN) tunneling and NN interactions. The
corresponding BH Hamiltonian reads:
\begin{eqnarray} \label{Ham}
 H &= \sum_{i;\sigma}\varepsilon(i,\sigma)
\hat{a}^\dagger_{i,\sigma}\hat{a}_{i,\sigma}-\sum_{\langle i,j
\rangle;\sigma}J_{1}\hat{a}^\dagger_{i,\sigma}\hat{a}_{j,\sigma}-\sum_{\langle
\langle i,j\rangle
\rangle;\sigma}J_{2}(i,\sigma)\hat{a}^\dagger_{i,\sigma}\hat{a}_{j,\sigma}
\\*\nonumber
 &+\sum_{i;\sigma}U(i,\sigma)\hat{a}^\dagger_{i,\sigma}\hat{a}^\dagger_{i,\sigma
}\hat{a}_{i,\sigma}\hat{a}_{i,\sigma}+\sum_{\langle i,j
\rangle;\sigma}U_{1}\hat{a}^\dagger_{i,\sigma}\hat{a}^\dagger_{j,\sigma}\hat{a}_
{i,\sigma}\hat{a}_{j,\sigma}\\*\nonumber
 &+\sum_{i}W\hat{a}^\dagger_{i,\uparrow}\hat{a}^\dagger_{i,\downarrow}\hat{a}_{i
,\uparrow}\hat{a}_{i,\downarrow}+\sum_{\langle i,j \rangle
}W_{1}(i)\hat{a}^\dagger_{i,\uparrow}\hat{a}^\dagger_{j,\downarrow}\hat{a}_{i,
\uparrow}\hat{a}_{j,\downarrow}.
\end{eqnarray}
In the Hamiltonian, $\hat{a}^{(\dagger)}_{i,\sigma}$ denotes the annihilation
(creation) operator of one boson of species $\sigma\in\{\uparrow,\downarrow\}$
at site $i$, and $\langle i,j \rangle$ ($\langle
\langle i,j\rangle \rangle$) indicates the summation over all the NN (NNN)
pairs.
The first three terms of the Hamiltonian in equation (\ref{Ham}) are
the on-site energy, NN hopping and NNN hopping terms, respectively, and the following
two terms are the intra-species on-site and NN interactions.
The last two terms describe the inter-species on-site and NN interactions.
Depending on the $i$-th site being deep or shallow for the $\sigma$
species, the on-site energy $\varepsilon(i,\sigma)$ vanishes for deep sites and 
assumes a value of $\varepsilon$ for shallow sites, and the NNN hopping matrix element
$J_{2}(i,\sigma)$ becomes $J_{2,d}$ or $J_{2,s}$, respectively. Similarly, there
are two in general different values $U_d$ and $U_s$ for the intra-species
on-site interaction
strength $U(i,\sigma)$. The NN inter-species interaction $W_1(i)$ takes the
values $W_{1,d}$ (neighboring atoms in their respective deep sites) and
$W_{1,s}$ (neighboring atoms in their respective shallow sites).
The NN hopping matrix element $J_1$, the NN intra-species interaction strength
$U_1$ and the on-site inter-species interaction strength $W$ are, however,
site- and spin-independent.
In the calculation of the amplitudes for the interaction processes, the
intra- and inter-species scattering lengths are taken to be equal.
Figure \ref{figure2} shows the parameter values we work with
as a function of the lattice depth $V_0$.

\begin{figure}
\centering
 \includegraphics[width=0.75\textwidth]
{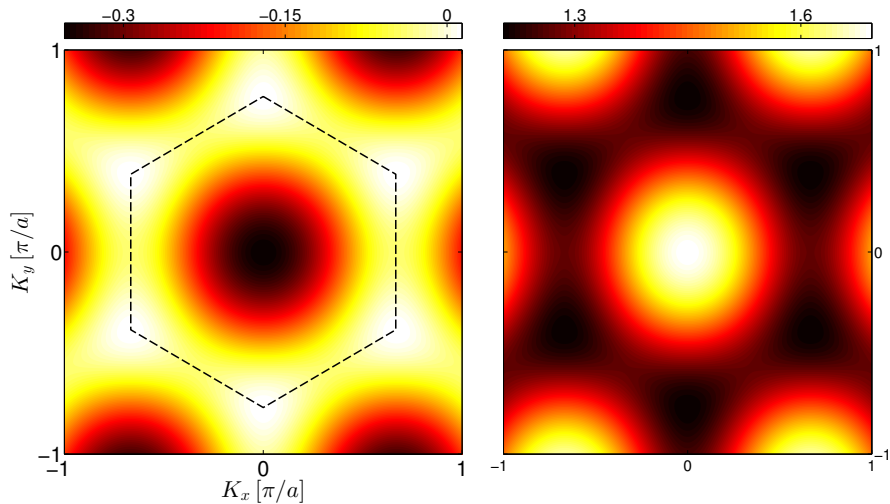}
\caption{The single particle eigenenergy
as a function of quasi-momenta for both the $\uparrow$ and
$\downarrow$ bosons, calculated at $V_0=1.5 E_R$. The left and right panels correspond to
the lower and upper branch, respectively.}
\label{figure3}
\end{figure}

Before turning to the many-body problem, let us briefly review the single
particle properties of this system. The two species share the same single
particle band structure, consisting of two branches as shown in figure
\ref{figure3}. In contrast to what is found for a standard graphene-type
honeycomb lattice, a gap between the two branches is opened due to the energy
offset between A and B sites.

\begin{figure}
\centering
 \includegraphics[width=0.45\textwidth]
{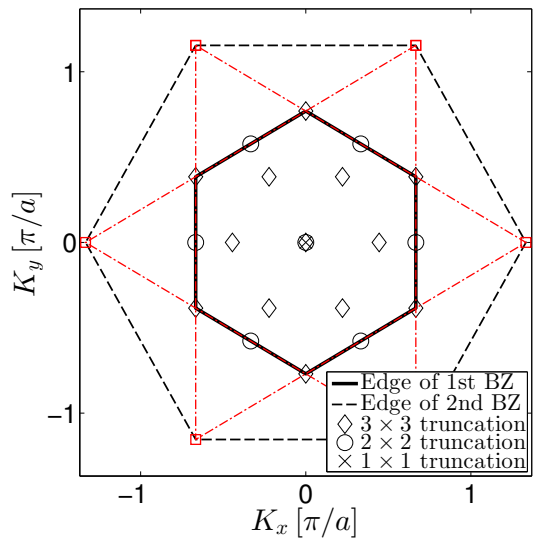}
\caption{Edges of the first (solid line) and the second (dashed line) Brillouin
zone in quasi-momentum space. Markers within the first Brillouin zone
indicate the quasi-momenta sampled by our different truncation schemes (for illustration, we also show the $3\times3$ case here).
The squares at the second Brillouin zone edges mark the first-order Bragg peaks, 
forming two equilateral triangles ($\triangleleft$ and $\triangleright$) shown as dashed-dotted lines.
}
\label{figure4}
\end{figure}

In our numerical simulations, we are limited to finite system sizes. 
The truncation
domains we focus on here are a single ($1 \times 1$) unit cell and $2\times2$ unit cells (cf. figure \ref{figure12}), 
although we have also tried larger domains where however convergence is hard to ensure. 
For all truncation schemes, periodic boundary conditions are imposed.
This results in an enhancement of the NN hopping strength $J_1$ by a
factor of three for the $1\times1$
truncation compensating the reduced coordination number. In principle, the
reduction
of the coordination number can affect also the NN interaction processes, which,
however, turn out to be negligibly small and are therefore not
renormalized. Furthermore, we remark that the NNN hopping
term is
neglected when truncating to a single unit cell or $2\times2$ unit cells to
avoid an artificial hopping from one site to itself. Finally, truncations in 
real space necessarily imply a corresponding discrete sampling of the continuous
quasi-momentum
space. In figure \ref{figure4} we illustrate the sets of discrete quasi-momenta
captured by our different real space truncation schemes.

\section{Method}\label{sec_method}
In this work, we apply ML-MCTDHB \cite{MLB1,MLB2} which is a highly
flexible {\it ab initio} method for solving the time-dependent many-body
Schrödinger equation for bosonic systems and in particular bosonic mixtures. The
method is based on expanding the total wave function with respect to a
time-dependent variationally optimized basis, which allows to capture the
important correlation effects using a numerically feasible basis size.
We employ imaginary time propagation in order to find the ground state of the
binary mixture in the SDHL potential by means of ML-MCTDHB.

The power and flexibility of ML-MCTDHB mainly lies in the multi-layer ansatz for
the total wave function. Going through this layer by layer, the total
wavefunction is first expanded as
\begin{eqnarray}\label{MLB_exp1}
 |\Psi(t)\rangle &=\sum_{i,j=1}^{M}A_{i,j}(t)|\psi^{\uparrow}_i
(t)\rangle|\psi^{\downarrow}_j(t) \rangle,
\end{eqnarray}
where the time-dependent basis for the
$\sigma$ species $\{|\psi^{\sigma}_i(t)\rangle\}|^M_{i=1}$ consists of the
so-called species-layer single particle functions (SPFs). These species         
layer SPFs are bosonic many-body states and as such expanded as
\begin{eqnarray}\label{MLB_exp2}
 |\psi^{\sigma}_i (t)\rangle&=\sum_{{\bf n}} B^{\sigma}_{i;\bf n}(t)| {\bf n}
;t\rangle_\sigma.
\end{eqnarray}
Here $|{\bf n};t \rangle_\sigma$ denotes the bosonic number state
obtained by populating the time-dependent particle layer SPFs
$\{|\varphi^{\sigma}_j (t)\rangle\}|^m_{j=1}$ according to the occupation
numbers ${\bf n}=(n_1,n_2,...,n_m)$ and the sum runs over all occupation
numbers that add up to the number of $\sigma$ bosons. In this work, we are only
interested in balanced mixtures of $N$ bosons per species.
Finally, the particle layer SPFs, which constitute the instantaneous truncated
basis within the single particle Hilbert space, are
expanded in terms of the static Wannier states
\begin{eqnarray}\label{MLB_exp3}
 |\varphi^{\sigma}_j(t)\rangle&=\sum_{l}C^{\sigma}_{j;l}(t)|l_\sigma\rangle
\end{eqnarray}
where $|l_\sigma\rangle$ denotes the Wannier state of species $\sigma$ in the
$l$-th site of the SDHL lattice, as defined in \ref{app_trafos}.
In this recursive
way the expansion coefficients $\{A_{i,j},B^{\sigma}_{i;{\bf
n}},C^{\sigma}_{i;j}\}$ determine the total wave function, and their time
evolution is governed by equations of motion \cite{MLB1,MLB2} obtained from a
time-dependent variational principle \cite{McLachlan,equiv_var_princp},
which ensures the optimal choice of the SPF basis on all layers.
Applying imaginary time propagation we thus obtain the expansion coefficients
for the ground state of the bosonic mixture.

The numerical control parameters of the above expansions are the numbers $M$ and
$m$ of species and particle layer SPFs of each species. Due to the symmetry
between the two species, these numbers are taken to be independent of $\sigma$.
In the limit of $M=m=1$, the ML-MCTDHB expansion reduces to the mean-field
ansatz, while the calculation becomes numerically exact
for $m$ equal to the number of lattice sites and $M=(N+m-1)!/[N!(m-1)!]$.
Summarizing, for a given system size we have to ensure
that our simulations are converged with respect to the numbers of
basis functions $(M,m)$ such that all relevant correlations are captured.

\section{Results}
Having determined the ground state many-body wave function
$|\Psi\rangle$ with the ML-MCTDHB method,
we need to establish a link to time-of-flight measurements. Assuming that
interactions do not
play any role during the expansion of the binary Bose mixture, an ideal
single shot absorption measurement after long time-of-flight effectively means
to measure
destructively
the joint $(N_\downarrow,N_\uparrow)$-particle free-space momentum distribution
{\it before} the expansion
\cite{Feshbach}.
In the following, we will focus on the first order Bragg peaks in the
time-of-flight images: As indicated in
figure \ref{figure4}, we group
these six free-space momenta into two classes, namely the set of momenta in the
triangle
pointing to the right (left)
denoted by $\triangleright=\{{\bf b}_1,{\bf b}_2,{\bf b}_3\}$
($\triangleleft=\{-{\bf b}_1,-{\bf b}_2,-{\bf
b}_3\}$). In \ref{app_trafos}, we show that all elements of
$\triangleright$ ($\triangleleft$)
are equivalent within our two-band approximation. Therefore,
we may single out two representatives named ${\bf r}\in\triangleright$ and
${\bf l}\in\triangleleft$. Moreover, we denote the zero free-space momentum
as ${\bf k}_0$.

In the following, we will first discuss our findings for the experimental
parameters of \cite{TSF_exp}, in order to quantify how
well this system is described by a mean-field approach.
Afterwards, we will consider potentially relevant inter-species correlation
effects, complementing the discussion of the experiment itself with some results
for detuning energies between the shallow and deep lattice sites which are
smaller than the experimentally implemented ones.

In both considerations, we  find that the reduced one-body free-space
momentum
density of the species $\sigma$,  $\rho^{\sigma}_1({\bf k})=\frac{1}{N_\sigma}
\langle\Psi|\hat\psi_{\sigma}^\dagger({\bf k})\hat\psi_{\sigma}({\bf
k})|\Psi\rangle$ with $\hat \psi_{\sigma}({\bf k})$ denoting the
annihilation operator of a free space momentum state ${\bf k}$, does not
feature TSF properties. In fact, we have not found any numerical hints that
the ground state $|\Psi\rangle$ might be degenerate. Therefore,
 $|\Psi\rangle$ features both a spin exchange and a six-fold rotation
symmetry implying
$\rho^{\downarrow}_1({\bf q})=\rho^{\uparrow}_1({\bf q})$ with
${\bf q}\in\{{\bf r},{\bf l}\}$ as well as
$\rho^{\sigma}_1({\bf r})=\rho^{\sigma}_1({\bf l})$ meaning that the average
of species-selective
histograms of $(N_\downarrow,N_\uparrow)$-particle
measurements over many shots does not give any TSF signatures.

\subsection{Ground state properties in the parameter regime of the experiment}
In this section we focus on the parameter values at which the experiment was performed.
As explained above, numerical simulations have been performed for different lattice truncations.
For each of these, the lattice depth was varied from weak to strong.
In the $1\times1$ truncation, we consider 3 $\uparrow$ and 3
$\downarrow$ bosons, corresponding to a filling factor of $3$ per unit cell
for each species. For the $2\times2$ truncation, we consider a filling
factor per unit cell being equal to unity for each species. Larger filling
factors are computationally prohibitive for the $2\times2$ due to the slow
convergence with respect to the number of SPFs on both layers as discussed in
section \ref{intercoll}.

\begin{figure}
\centering
 \includegraphics[width=0.8\textwidth]
{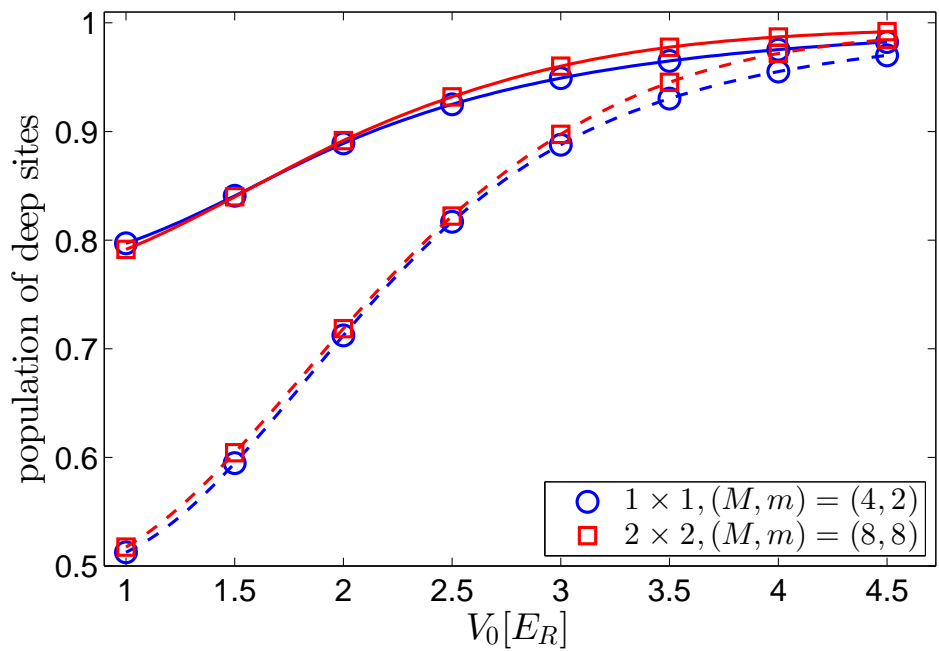}
\caption{Population of $\uparrow$ bosons in their deep sites as a function of the lattice depth $V_0$,
calculated within the $1\times1$ (blue lines) and $2\times2$ (red lines)
truncations. The solid lines are calculated with the experimental parameters
shown in figure \ref{figure3}, and
the dashed lines are calculated for a smaller
energy offset $\varepsilon^*$ being discussed in section \ref{intercoll}.
}
\label{spatial_occu}
\end{figure}

We start by discussing our results for the one-body density in position space.
In figure \ref{spatial_occu} we present the population of $\uparrow$ bosons in their corresponding deep sites, $\rho_{A}(\uparrow)$, as a function of the optical lattice depth. Due to the spin
symmetry, the population of $\downarrow$ bosons is given by
$\rho_A(\uparrow)=\rho_B(\downarrow)$, while the $\sigma$ species occupations
at sites $A$ and $B$ add up to unity:
$\rho_A(\sigma)+\rho_B(\sigma)=1$.
It can be seen that the $\uparrow$ bosons already show
a tendency to localize in the $A$ sites in the weak lattice depth regime,
and the localization becomes stronger as the lattice depth increases, following the increase of the state-dependent on-site energy offset $\varepsilon$.
As the $\downarrow$ bosons show the same tendency to be localized in their deep sites, 
$i.e.$ the $B$ sites, the spatial overlap between $\uparrow$ and $\downarrow$ bosons becomes vanishingly small for large V0.
As a consequence, increasing $V_0$ leads to a suppression of inter-species correlations, as will be discussed below.

To quantify how close the system is to a mean-field state, we further calculate
so-called depletions for various subsystems: Firstly focussing on
a $\sigma$ boson, the corresponding reduced one-body density operator is
given by the partial trace of the total system state over all atoms but
a single one of spin $\sigma$: $\hat \rho_1^\sigma={\rm
Tr}_{\sigma; {\rm part}    }|\Psi\rangle\!\langle\Psi|$. Denoting the largest eigenvalue
of $\hat \rho_1^\sigma$ as $\lambda_p^\sigma$, the depletion of a $\sigma$
atom is defined as $d_\sigma=1-\lambda_p^\sigma$, which turns out to be
independent of $\sigma$ due to the spin-exchange symmetry, $d_\sigma\equiv d$.
According to
\cite{Onsager_Penrose_BEC_liquid_He_PR_1956}, $d$ measures the degree of
condensation of the $\sigma$ bosons: $d$ close to zero implies that
intra-species correlations are negligible, i.e. all the $\sigma$ bosons
essentially
occupy the same single particle state. In this way, $d\approx 0$ means that a
mean-field ansatz for the state of the $\sigma$ species is applicable, i.e.
that one may set $m=1$ in the wave function ansatz
(\ref{MLB_exp1},\ref{MLB_exp2}).
Conversely, a significant value of $d$
($<1$) indicates strong correlations between a $\sigma$ boson and all the
other atoms.

This concept can be transferred to the whole $\sigma$ species whose reduced
density operator is given by the partial trace of the total system state over
the atoms of opposite spin: $\hat \rho_{\rm spec}^\sigma={\rm
Tr}_{\sigma; {\rm spec}    }|\Psi\rangle\!\langle\Psi|$. With $\lambda_s^\sigma$ as
its largest eigenvalue, we similarly define the depletion on the species
layer as $D=1-\lambda_s^\sigma$ (which is again independent of the $\sigma$).
This depletion measures how strongly the total state deviates from a single
tensor product of a pure state for the $\uparrow$ species and a
 pure state for the $\downarrow$ species. For $D=0$, this factorization holds
exactly implying the validity of a mean-field ansatz ($M=1$) on the species
layer (\ref{MLB_exp1}). Significant values of $D$ ($<1$) imply the presence of
inter-species correlations resulting in a non-factorizable total wave function.
We remark that the depletions $d$ and $D$ measure
correlations in the system basis-independently.

\begin{figure}
\centering
 \includegraphics[width=0.7\textwidth]
{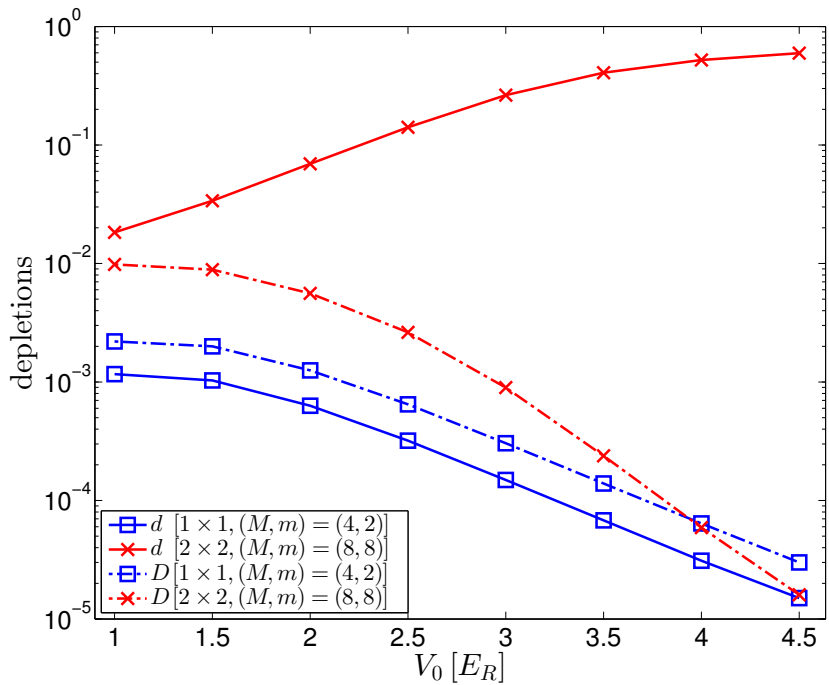}
\caption{Species-level and particle-level depletions $D$ (dashed-dotted lines) and $d$
(solid lines) as a function of the lattice depth. The colours denote the results from
different spatial truncations, with the corresponding $(M,m)$ given in the legend.
}
\label{depletion}
\end{figure}
Figure \ref{depletion} shows the numerically obtained depletions $D$ and $d$ for our different spatial truncation schemes.
Throughout, we find that the depletion $D$ is very weak (below one percent) for all lattice depths considered, decreasing further for increasing $V_0$.
This latter trend is in agreement with the above discussion of the inter-species correlations (which are small throughout) being further suppressed due to a spatial separation of the $\uparrow$ and $\downarrow$ bosons.
Turning to the depletion $d$, we find a qualitative difference between the $1\times 1$ and $2\times 2$ truncation schemes.
While in the effectively one-dimensional $1 \times 1$ simulations, $d$ is on the level of per mil or below, it assumes larger values in the $2 \times 2$ setup, starting from few percent for low $V_0$ and going up to around $0.5$ for the largest lattice depth we consider, indicating fragmentation of the system on the level of a single species for deep lattices.
This finding is consistent with approaching a Mott insulator phase for increasing $V_0$, with the bosons of a species fully localizing in their respective deep sites. In the simulation with a single unit cell the fragmentation accompanying the localization cannot be captured since only a single deep site is available for each species.\\
Overall, we observe that in the regime of small $V_0$ (where in the experiment the TSF was observed) there is no numerical evidence of sizable depletions. This indicates that a mean-field treatment of the system is indeed expected to give appropriate lowest order results, to which then quantum corrections apply. In the following we will turn to a detailed inspection of these beyond mean-field corrections, and their possible relation to the TSF signatures.

\subsection{Interspecies correlation effects} \label{intercoll}
In this section, we investigate inter-species correlation effects that
potentially could give rise to properties characteristic for the TSF.
From the discussion in the previous section it is generally expected that
such correlations are small for the experimental parameter values of
\cite{TSF_exp}, but we will illustrate that they are enhanced in other regions
of the parameter space.

One of the simplest quantities which may be sensitive to beyond mean-field signatures of the TSF properties is the inter-species
free-space momentum $g_2$-correlation function:
\begin{equation}\label{2b_inter_correlation_fct}
 g_2^{\downarrow,\uparrow}({\bf q}_1,{\bf q}_2) = \frac{1}{N_\downarrow N_\uparrow}
 \frac{\langle\Psi|
 \hat\psi_{\downarrow}^\dagger({\bf q}_1)
 \hat\psi_{\uparrow}^\dagger({\bf q}_2)
 \hat\psi_{\downarrow}({\bf q}_1)
 \hat\psi_{\uparrow}({\bf q}_2)|\Psi\rangle}
 {\rho^{\downarrow}_1({\bf q}_1)
 \rho^{\uparrow}_1({\bf q}_2)}.
\end{equation}

\begin{figure}
\setcounter{subfigure}{0}
 \centering
\subfigure{
\centering
\includegraphics[width=0.5\textwidth]
{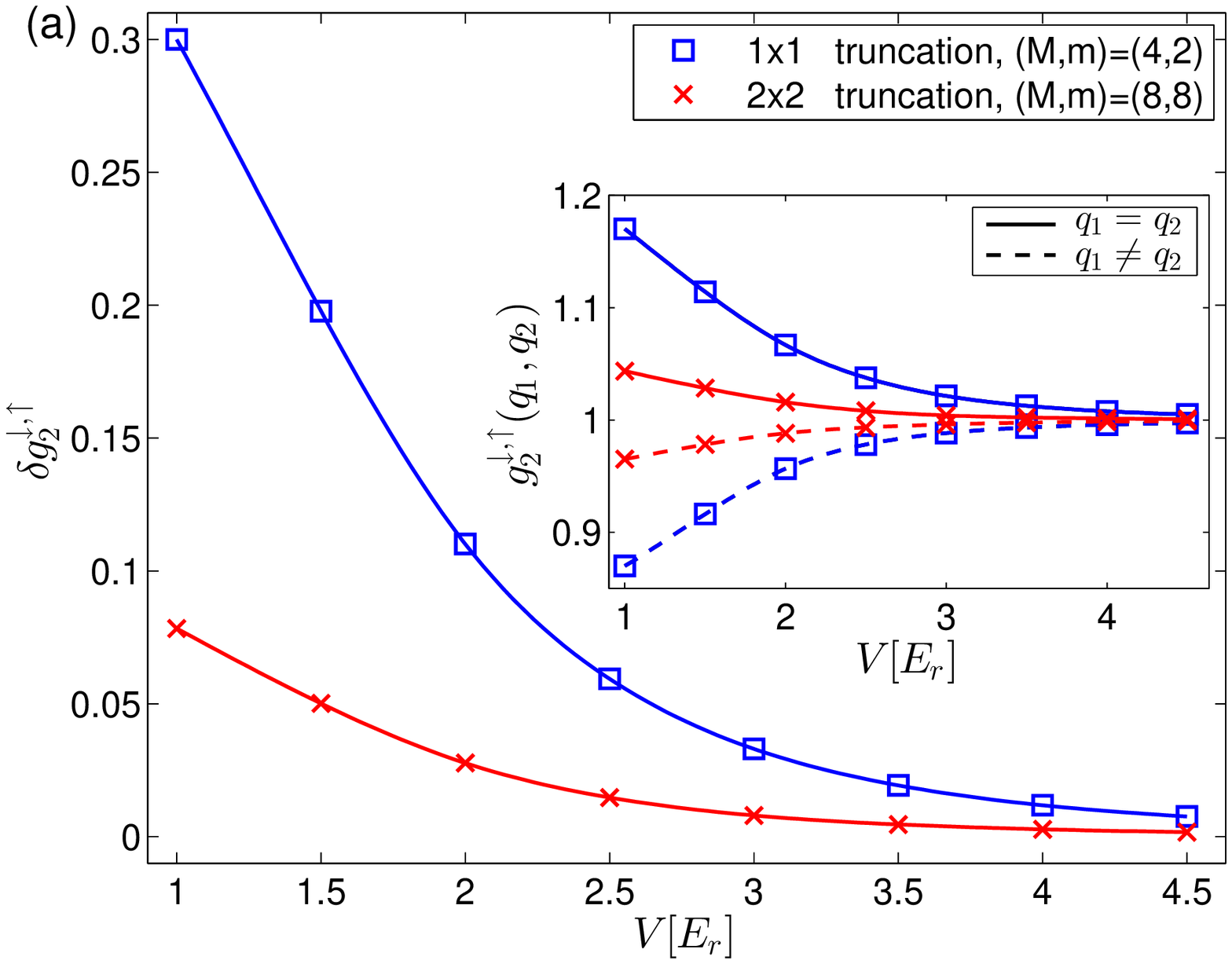}
}
\subfigure{
 \centering
 \includegraphics[width=0.5\textwidth]
 {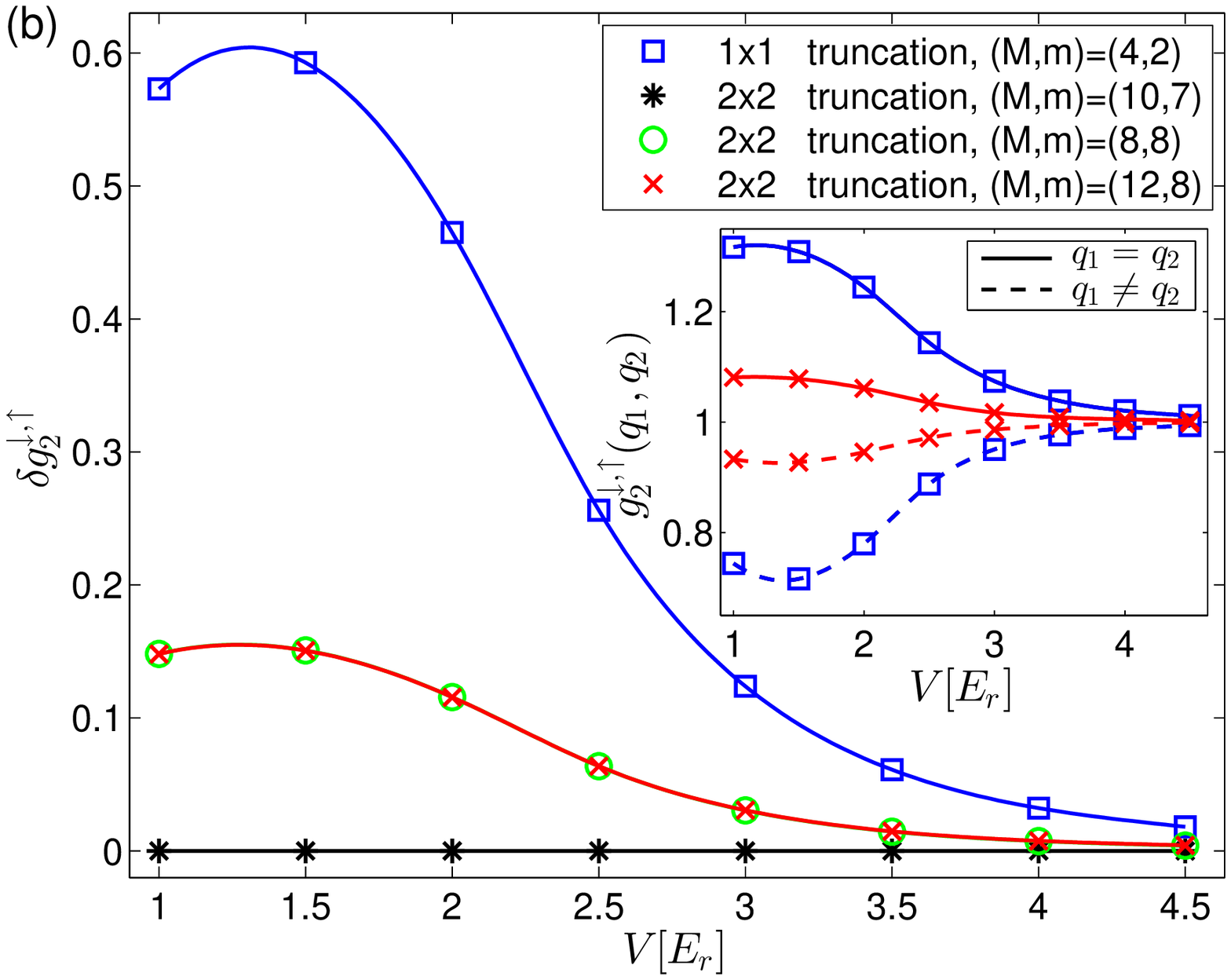}
}
\caption{Inter-species correlation difference $\delta
g^{\downarrow\uparrow}_2$
as a function of lattice depth, for (a) 
the experimental value of the energy offset as shown in figure \ref{figure2}
and (b) the by one recoil energy reduced $\varepsilon^*$. In the upper panel,
the
blue and red lines are for $1\times1$ and $2\times2$ truncations,
respectively, with the corresponding $(M,m)$ given in the legend.
In the lower panel, results for $2\times2$ truncation with different
$(M,m)$ are also given.
The insets in both panels depict
$g_2^{\downarrow,\uparrow}({\bf q}_2,{\bf q}_1)$
with ${\bf q}_2={\bf q}_1$ (solid line) and ${\bf q}_2\neq{\bf q}_1$
(dashed line).
}
\label{deltag2}
\end{figure}

We note the spin exchange symmetry $g_2^{\downarrow,\uparrow}({\bf q}_2,{\bf
q}_1)
=g_2^{\downarrow,\uparrow}({\bf q}_1,{\bf q}_2)$
as well as the rotation symmetry $g_2^{\downarrow,\uparrow}({\bf r},{\bf r})
=g_2^{\downarrow,\uparrow}({\bf l},{\bf l})$. While
$g_2^{\downarrow,\uparrow}({\bf q}_1,{\bf q}_2)$ equals unity in
the mean-field approximation, correlations can result in inter-species bunching
($g_2^{\downarrow,\uparrow}({\bf q}_1,{\bf q}_2)>1)$ or
antibunching ($g_2^{\downarrow,\uparrow}({\bf q}_1,{\bf q}_2)<1)$, which
indicates that the spin exchange or six-fold rotation symmetry is
stochastically violated in a {\it correlated manner} when detecting two bosons
of opposite spin: If for instance $g_2^{\downarrow,\uparrow}({\bf
r},{\bf r})$
turns out to be larger than unity, we may conclude that, given one spin
$\uparrow$ boson
has been destructively detected in one triangle, the probability for
subsequently detecting
a spin $\downarrow$ boson in the very same triangle is enhanced compared to
statistical
independence. Moreover, we compare these
correlations by introducing:
\begin{equation}\label{2b_inter_correlation_diff}
 \delta g_2^{\downarrow,\uparrow} = g_2^{\downarrow,\uparrow}({\bf r},{\bf r})-
                    g_2^{\downarrow,\uparrow}({\bf r},{\bf l}).
\end{equation}
Finding $\delta g_2^{\downarrow,\uparrow}<0$ could consequently be interpreted
as a manifestation of (P2) (see introduction) on the level of two-body measurements, i.e.
that two bosons of
different spin are more likely to be found in different triangles than in
the same one.

Figure \ref{deltag2} depicts $\delta g^{\downarrow\uparrow}_2$ and, as an inset,
also the individual addends $g_2^{\downarrow,\uparrow}({\bf r},{\bf r})$ and
$g_2^{\downarrow,\uparrow}({\bf r},{\bf l})$ as a function of the lattice depth
for various truncation sizes. We show the quantities for the experimental
parameters, which as expected are small since the system is in the mean field
regime. For the sake of illustration, we have also included results for the
on-site energy detuning $\varepsilon^*=\varepsilon-E_R$, which is reduced by one
recoil energy compared to the value of the
experiment \cite{TSF_exp} while all other
parameters in the Hamiltonian are kept fixed. It can be observed that the
inter-species correlations are enhanced for the reduced detuning
$\varepsilon^*$ in comparison to the results for $\varepsilon$.
Understandably, reducing the
energy difference between the species-dependent deep and shallow sites reduces
the bosons' tendency to separate into the sublattices. In turn, the
inter-species spatial overlap is larger, enabling the emergence of
correlations. Enhancement of the correlations is also
observed when increasing the inter-species interaction (results not shown).

Notably, the values of both $|\delta g^{\downarrow\uparrow}_2|$
and $|g_2^{\downarrow,\uparrow}({\bf q}_1,{\bf q}_2)-1|$ are largest for the
single unit cell truncation where the simulated system is effectively
one-dimensional, while they tend to decrease when going to larger truncation
sizes where the true two-dimensional structure of the extended lattice is
correctly taken into account. The
magnitude of correlations
is reduced for larger truncation sizes as more energetically favourable
states are
included in the discrete sampling of the quasi-momentum space: As shown in
\ref{app_g2_coher}, a non-vanishing
$\delta g_2^{\downarrow,\uparrow}$ requires a finite coherence between
the single particle ground and highest excited state within the single band approximation.
If now
energetically lower lying states become available when increasing the
truncation size, atoms generically tend to occupy these states and thereby
reduce the inter-branch coherences related to $\delta
g_2^{\downarrow,\uparrow}$.

Importantly, for the two different values of the energy offset,
we only observe positive values
of $\delta
g^{\downarrow\uparrow}_2$, indicating that due to inter-species
correlations one
$\uparrow$ and one $\downarrow$ boson slightly tend to concentrate in the
same triangle, in contrast to (P2). Thus, even if the weak inter-species
correlations of
this kind are taken into consideration (and as we have seen they may become more
relevant in regions of the parameter space away from the experimental values) we
find no hint that they can cause the emergence of a (P2) signature.

We furthermore remark that the correlation measure $\delta
g_2^{\downarrow,\uparrow}$ converges slowly with the number of SPFs: It can be
inferred from figure \ref{deltag2} that one has to provide as many particle
layer SPFs as there are sites, which corresponds to the numerically
exact limit with respect to the particle layer basis
and stems from the fact that coherences between the energetically most
separated single particle eigenstates must be resolved for having $\delta
g_2^{\downarrow,\uparrow}$ non-vanishing (cf. \ref{app_g2_coher}).

So far, we have only discussed two-body correlation properties.
In some situations, however, it can be difficult to generalize the results for
two-body
observables to $N$ particle measurements (cf. e.g. the
discussion \cite{quantum_entangled_dark_solitons_in_optical_lattices_Carr_PRL09,
comment_on_Carr_PRL_PRA_Sacha_PRL2010,
reply_on_comment_on_Carr_PRL_PRA_Carr_PRL2010,
delande_many-body_2014}). For this reasons, we evaluate here the joint
probability distribution for detecting
$n_0^\sigma$, $n_\triangleleft^\sigma$, $n_\triangleright^\sigma$
bosons of spin $\sigma$ at zero free-space momentum, in $\triangleleft$, in
$\triangleright$, respectively:
$P(n_0^\uparrow,n_\triangleleft^\uparrow,n_\triangleright^\uparrow;
n_0^\downarrow,n_\triangleleft^\downarrow,n_\triangleright^\downarrow)$.
Restricting ourselves to the $1\times 1$ truncation and bearing in mind that
the states within $\triangleleft$ ($\triangleright$) are equivalent
within the lowest band approximation (cf. \ref{app_trafos}), this
probability distribution is given by:
\begin{eqnarray}
 P(n_0^\uparrow,n_\triangleleft^\uparrow,n_\triangleright^\uparrow;
n_0^\downarrow,n_\triangleleft^\downarrow,n_\triangleright^\downarrow) &=&
\frac{1}{16}\prod_{\sigma\in\{\uparrow,\downarrow\}}
(n_\triangleleft^\sigma+1)
(n_\triangleleft^\sigma+2)(n_\triangleright^\sigma+1)(n_\triangleright^\sigma+2)
\\*\nonumber
&&\times
\tilde P(n_0^\uparrow,n_\triangleleft^\uparrow,n_\triangleright^\uparrow;
n_0^\downarrow,n_\triangleleft^\downarrow,n_\triangleright^\downarrow).
\end{eqnarray}
Here, $\tilde P(n_0^\uparrow,n_{\bf l}^\uparrow,n_{\bf r}^\uparrow;
n_0^\downarrow,n_{\bf l}^\downarrow,n_{\bf r}^\downarrow)$ refers to the
probability for finding $n_0^\sigma$, $n_{\bf l}^\sigma$,
$n_{\bf r}^\sigma$ bosons of spin $\sigma$ in the free-space momentum states
${\bf k}_0$, ${\bf l}$, ${\bf r}$, respectively, which can be calculated in
terms of an expectation value of the operators
 $\hat \psi^{(\dagger)}_{\sigma}({\bf k})$ (cf.
(\ref{field_op_trafo}) for ${\bf k}={\bf l},{\bf r}$).

\begin{figure}
 \centering
 \includegraphics[width=0.8\textwidth]
{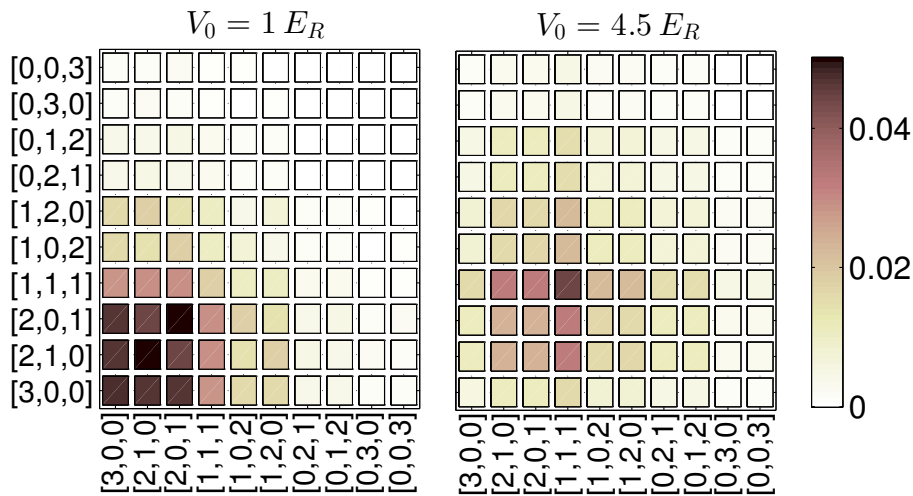}
\caption{The joint probability for finding
$[n_0^\uparrow,n^\uparrow_\triangleright,n^\uparrow_\triangleleft]$
bosons of spin $\uparrow$ (x-axis) and
$[n_0^\downarrow,n^\downarrow_\triangleright,n^\downarrow_\triangleleft]$
 bosons of spin $\downarrow$ (y axis) in ${\bf k}_0$, $\triangleright$,
$\triangleleft$,
respectively,
for the energy offset $\varepsilon^*$ and $V_0=1 E_R$
(left) as well as $V_0=4.5 E_R$ (right).
}
\label{wfproj}
\end{figure}

In figure \ref{wfproj}, the probability distribution
$P(n_0^\uparrow,n_\triangleleft^\uparrow,n_\triangleright^\uparrow;
n_0^\downarrow,n_\triangleleft^\downarrow,n_\triangleright^\downarrow)$ is
depicted for the reduced energy offset $\varepsilon^*$ and $V_0=1E_R$ as well as
$V_0=4.5E_R$. We clearly see that
detection events with two or three bosons per species in ${\bf
k}_0$ dominate in probability for the shallow lattice while for the deep lattice
the probability is largest for detecting one $\sigma$ boson in ${\bf
k}_0$, another one in $\triangleleft$ and the third one in $\triangleright$.
Moreover,
we observe for the deep lattice that the four combinations of finding
two $\uparrow$ and two $\downarrow$ bosons in ${\bf k}_0$ and the remaining
$\uparrow$ and $\downarrow$ boson in $\triangleleft$ / $\triangleright$
are of the same probability indicating the absence of inter-species two-body
correlations, i.e. $g_2^{\downarrow,\uparrow}({\bf q}_1,{\bf q}_2)\approx 1$
for ${\bf q}_i\in\{{\bf r},{\bf l}\}$. In contrast to this,
the presence of inter-species correlations can be witnessed for the shallow
lattice: The largest probability
$P(2,1,0;2,1,0)=P(2,0,1;2,0,1)$ relates to detecting two bosons per species
in ${\bf k}_0$ and the remaining two bosons of opposite spin in the very same
triangle, which is consistent with $\delta
g^{\downarrow\uparrow}_2>0$ indicating the bunching of bosons of opposite
spin in the same triangle. Therefore, also the joint probability distribution
gives a slight inter-species bunching effect opposite to (P2).
In view of (P1), we have also considered the marginal probability distribution
of
finding $n_0^\uparrow$, $n_\triangleleft^\uparrow$, $n_\triangleright^\uparrow$
bosons of spin $\uparrow$ at zero free-space momentum, in $\triangleleft$, in
$\triangleright$ - irrespectively of the state in which the $\downarrow$ bosons
are found (plot not shown). Both this marginal probability distribution
and the full probability distribution indicate that the first order Bragg peak
triangles are most likely either unoccupied or occupied by a single boson.
This finding can in principle be related to an asymmetric occupation of the
triangles within one species, i.e. (P1). However, we cannot prove this tendency
as this would require to consider more than three bosons per species, i.e.
going at least to the $2\times2$ truncation.
\section{Conclusions and discussion}
We have studied the ground state properties of a binary bosonic mixture confined in a state-dependent honeycomb lattice
for different values of the lattice depth.
Our results show that in the parameter regime related to the experimental setup in \cite{TSF_exp}, the system can be
relatively well described by a mean-field approximation, and only small corrections in the form of inter-species quantum correlations are observed.
No indications of a ground state degeneracy are found.
We study possible relations of the correlation effects beyond mean-field to potential signatures of the TSF and find that the correlations can indeed 
modify the population of the first order Bragg peaks after time-of-flight. The specific properties characteristic of the twisted superfluid are however not observed in our simulations.
Our results imply that the emergence of the twisted superfluid cannot be explained by taking into account quantum corrections within the lowest-band extended Bose-Hubbard Hamiltonian.
The most severe restrictions of our investigations are the finite system size, limiting us to a regime far away from an infinitely extended lattice.
This truncation in real space implies a discrete selection of quasi-momenta as illustrated in figure \ref{figure4}, which does not capture the long wavelength excitations of the system.
Given the slow convergence of the correlation properties with respect to the number of basis functions $(M,m)$, going to larger system sizes is computationally prohibitive.
Furthermore, our model Hamiltonian neglects higher band effects whose importance for the twisted superfluid effect therefore remains as an open question.
One may also speculate that neglected processes in the lowest-band extended BH model, such as the density-induced tunneling or pair tunneling terms, may be of importance in the emergence of the twisted superfluid.

\ack
P.S., K.S. and L.C. gratefully acknowledge funding by the
Deutsche Forschungsgemeinschaft in the framework of the SFB 925 ``Light induced
dynamics and control of correlated quantum systems''. S.K. and
J.S. gratefully acknowledge a scholarship by the Studienstiftung
des deutschen Volkes.
\appendix
\section{Transformation of single particle basis}\label{app_trafos}
\setcounter{section}{1}

Due to the local character of the BH Hamiltonian (\ref{Ham}), our simulations
are performed in real space. Afterwards the result has to be transformed into
the free space momentum basis in order to establish a link to the time-of-flight measurements. 
Moreover, one can relate certain inter-species correlations in
the free momentum space to coherences between quasi-momentum states as shown
in \ref{app_g2_coher}. For these reasons,
we provide the transformations between the Wannier, quasi-momentum
and free-space momentum bases, which are relevant for this work.

Let $w^\sigma_{A/B}(\bf x)$ denote
the Wannier function of species $\sigma$ for a site of type A/B centered
at position $\bf x=0$ as constructed in
\cite{Wannier}. Then the Wannier basis $\{|{\bf R}\rangle_{\sigma,\rm A},
|{\bf R}\rangle_{\sigma,\rm B}\}$\footnote{Note that in the main text 
the compact notation $|j \rangle_\sigma$ with a joint site index $j$ has
been employed.} can be obtained by Bravais translations
\begin{eqnarray}\label{Wannier_basis}
 \langle {\bf x}|{\bf R}\rangle_{\sigma,\rm A}=w^\sigma_{\rm A}({\bf x}-{\bf
R}),\quad \langle {\bf x}|{\bf R}\rangle_{\sigma,\rm B}=w^\sigma_{\rm B}({\bf
x}-{\bf R } -\bf r) .
\end{eqnarray}
Here, ${\bf R}=n_1{\bf R}_1+n_2{\bf R}_2$ with $n_1,n_2\in \mathbb{Z}$ refers to
a site of the A sublattice,
and the Bravais vectors of this hexagonal lattice are given by ${\bf
R}_1=\lambda(\frac{1}{\sqrt{3}},\frac{1}{3})$ and ${\bf
R}_2=\lambda(0,\frac{2}{3})$.
The relative distance between an A and a B site within a unit cell reads ${\bf
r}=\lambda(-\frac{\sqrt{3}}{9},-\frac{1}{3})$, with norm $|{\bf r}| = a$. We note the relation between the
Bravais and the reciprocal lattice vectors 
${\bf R}_i \cdot {\bf b}_j=2\pi\delta_{ij}$ for $i=1,2$.

Projecting the Wannier basis onto free-space momentum states, i.e.
plane waves $\langle{\bf x}|{\bf k}\rangle \propto\exp(i{\bf kx})$, provides
the following matrix for the transformation from the Wannier to the
free-momentum basis:
\begin{equation}\label{Wannier_free_space_mom_trafo}
\langle{\bf k}|{\bf R}\rangle_{\sigma,\rm A}\propto e^{-i{\bf kR}}
\phi^\sigma_{\rm A}({\bf k}), \quad \langle{\bf k}|{\bf R}\rangle_{\sigma,\rm
B}\propto e^{-i{\bf k(R-r)}} \phi^\sigma_{\rm B}({\bf k})
\end{equation}
where we have defined the Fourier transformation of $w^\sigma_{\rm A/B}(\bf x)$
as
\begin{equation}
  \phi^\sigma_{\rm A/B}({\bf k}) = \frac{1}{2\pi}\int{\rm d}^2x\,w^\sigma_{\rm
A/B}(\bf x) e^{-i{\bf kx}}.
\end{equation}
We note that $\langle{\bf k}|{\bf R}\rangle_{\sigma,A}$ and
$\langle{\bf k}|{\bf R}\rangle_{\sigma,B}$ feature a phase factor difference of
$\exp(-i{\bf kr})$, due to the finite separation
of A and B sites within the same unit cell. Relation
(\ref{Wannier_free_space_mom_trafo}) can be used for transforming
the quasi-momentum states $|{\bf Q}\rangle_{\sigma,\rm A/B}$ (which we label 
with capital letters ${\bf Q}$ to distinguish them from free-space momentum
states):
\begin{equation}\label{quasi_momentum_states}
|{\bf Q}\rangle_{\sigma,\rm A/B}\propto \sum_{{\bf R}} e^{i{\bf Q R}}|{\bf
R}\rangle_{\sigma,\rm A/B},
\end{equation} 
to free-space momentum states $|{\bf k}\rangle$.
Projecting onto $|{\bf k}\rangle$ and employing plane wave orthogonality results
in:
\begin{equation}\label{trafo_quasi_free_space_1}
 \langle {\bf k}|{\bf Q}\rangle_{\sigma,\rm A}\propto \phi^\sigma_{\rm A}({\bf
k})\sum_{{\bf R}}  e^{i{\bf (Q-k) R}} \propto \phi^\sigma_{\rm A}({\bf k})
\delta_{\mathcal Q(\bf k), \bf Q}
\end{equation}
where we have introduced the map $\mathcal Q$ that projects its argument vector
into the first Brillouin zone, i.e. shifts it by an appropriate reciprocal
lattice vector. Correspondingly, we have $\langle {\bf k}|{\bf
Q}\rangle_{\sigma,\rm B}\propto \exp(-i{\bf kr}) \phi^\sigma_{\rm B}({\bf k})
\delta_{\mathcal Q(\bf k), \bf Q}$.

Using these relations together with the orthonormality and assumed completeness
of the lowest band Bloch basis, we find the following representation of free
momentum states in terms of the quasi-momentum basis:
\begin{equation}\label{quasi_mom_to_free_space_trafo}
 |{\bf k}\rangle_\sigma \simeq \mathcal{N} \left[\phi^\sigma_{\rm A}({\bf
k})^*|\mathcal Q ({\bf k})\rangle_{\sigma,\rm A} 
 +e^{i{\bf kr}}\phi^\sigma_{\rm B}({\bf k}) ^*|\mathcal Q ({\bf
k})\rangle_{\sigma, \rm B}\right],
\end{equation}
where $\mathcal{N}$ is a normalization constant. In this expression, we have
explicitly introduced the species index on the left hand side.
We emphasize that the representation (\ref{quasi_mom_to_free_space_trafo}) is
an identity only within the truncation to the lowest band and
should be read as follows: $|{\bf k}\rangle_\sigma$ is given by the
expression on the right hand side plus a remainder which, however, is orthogonal
to our truncated Hilbert space and as such can be neglected when evaluating the
simulation results. In particular, we note that
(\ref{quasi_mom_to_free_space_trafo}) is not in conflict with e.g. the
orthogonality of free-space momentum states. 

Since the first order Bragg peaks in the long time-of-flight
absorption measurements are particularly relevant in view of the TSF properties, we concretise the transformation
(\ref{quasi_mom_to_free_space_trafo})
for these free-space momenta here.
The Wannier functions $w^\sigma_{\rm A/B}$ are real functions with three-fold
rotational symmetry and possess the symmetry $w^\uparrow_{\rm A/B}({\bf
x})=w^\downarrow_{\rm B/A}(-{\bf x})$ under spin exchange,
which greatly reduces the number of independent Fourier coefficients. In
particular we find the identities for $i,j,k,l\in\{1,2\}$:
\begin{eqnarray}
 \phi^\uparrow_{\rm A}({\bf b}_i)&=\phi^\uparrow_{\rm A}(-{\bf
b}_j)^*=\phi^{\downarrow}_{\rm B}({\bf b}_k)^*=\phi^\downarrow_{\rm B}(-{\bf
b}_l) =: \phi_{\rm A} ,\\ \nonumber
 \phi^\uparrow_{\rm B}({\bf b}_i)&=\phi^\uparrow_{\rm B}(-{\bf
b}_j)^*=\phi^\downarrow_{\rm A} ({\bf b}_k)^*=\phi^\downarrow_{\rm A}(-{\bf
b}_l) =: \phi_{\rm B}.
\end{eqnarray}
Thus, we can restrict to arbitrary representatives ${\bf r} \in \triangleright$
and ${\bf l} \in \triangleleft$ and find using
(\ref{quasi_mom_to_free_space_trafo}): 
 \begin{eqnarray}
 |{\bf r}\rangle_{\uparrow}&\simeq \tilde \mathcal{N}\left(\phi_{\rm A}^*|{\bf
Q=0}\rangle_{\uparrow, \rm A}+\phi^*_{\rm B}e^{-2i\pi/3}|{\bf
Q=0}\rangle_{\uparrow, \rm B}\right), \\*\nonumber
 |{\bf l}\rangle_{\uparrow}&\simeq \tilde \mathcal{N}\left(\phi_{\rm A} |{\bf
Q=0}\rangle_{\uparrow, \rm A}+\phi_{\rm B}e^{2i\pi/3}|{\bf
Q=0}\rangle_{\uparrow, \rm B}\right),\\*\nonumber
 |{\bf r}\rangle_{\downarrow}&\simeq \tilde \mathcal{N}\left(\phi_{\rm B}|{\bf
Q=0}\rangle_{\downarrow, \rm A}+\phi_{\rm A}e^{-2i\pi/3}|{\bf
Q=0}\rangle_{\downarrow, \rm B}\right),\\*\nonumber
 |{\bf l}\rangle_{\downarrow}&\simeq \tilde \mathcal{N} \left(\phi^*_{\rm B}|{\bf
Q=0}\rangle_{\downarrow, \rm A}+\phi^*_{\rm A}e^{2i\pi/3}|{\bf
Q=0}\rangle_{\downarrow, \rm B}\right),\\*\nonumber
\label{trafo_quasi_free_space_2}
\end{eqnarray}
where $\tilde \mathcal{N}$ denotes a normalization constant and the $\simeq$ symbol
has to be understood as in (\ref{quasi_mom_to_free_space_trafo}).

\section{Free-space momentum correlations and quasi-momentum coherences}
\label{app_g2_coher}
\setcounter{section}{2}
In this appendix, we establish a link between the correlation function
difference
$\delta g_2^{\downarrow,\uparrow}$ and inter-branch coherences given a
real-valued
total wave function in quasi-momentum space: By means of the transformation
(\ref{trafo_quasi_free_space_2}), we may express
the annihilation operator for a $\sigma$ boson in the free space momentum
${\bf q}\in\{{\bf r},{\bf l}\}$ as
\begin{equation}\label{field_op_trafo}
 \hat \psi_{\sigma}({\bf q}) \simeq {_{\;\sigma}\langle}{\bf q}|{\bf
Q=0}\rangle_{\sigma,A}
 \;\hat \alpha_\sigma + {_{\;\sigma}\langle}{\bf q}|{\bf Q=0}\rangle_{\sigma,B}
 \;\hat \beta_\sigma,
\end{equation}
where $\hat \alpha_\sigma$,
$\hat \beta_\sigma$ refer to the annihilation operators corresponding to
$|{\bf Q=0}\rangle_{\sigma,A}$, $|{\bf Q=0}\rangle_{\sigma,B}$, respectively.
The ground state being non-degenerate (as observed numerically) implies
$\langle \hat \psi^\dagger_{\uparrow}({\bf q}) \hat \psi_{\uparrow}({\bf q})
\rangle = \langle \hat \psi^\dagger_{\downarrow}({\bf q}) \hat
\psi_{\downarrow}({\bf q})
\rangle \equiv \bar{n}$. Introducing $\Theta=\phi_A^*\phi_B e^{2i\pi/3}$, we
find:
\begin{eqnarray}
\bar{n}^2\,\delta g_2^{\downarrow,\uparrow} &=&
 \langle
  \hat \psi^\dagger_{\downarrow}({\bf r}) \hat \psi_{\downarrow}({\bf r})
  \,\big[\,\hat \psi^\dagger_{\uparrow}({\bf r}) \hat \psi_{\uparrow}({\bf r})
  - \hat \psi^\dagger_{\uparrow}({\bf l}) \hat \psi_{\uparrow}({\bf l})
\,\big]\,
  \rangle\\*\nonumber
  &=& 2i\tilde\mathcal{N}^2{\rm Im}(\Theta)\;
  \langle
  \hat \psi^\dagger_{\downarrow}({\bf r}) \hat \psi_{\downarrow}({\bf r})
  \,\big[\,\hat \alpha^\dagger_\uparrow \hat \beta_\uparrow -
   \hat \beta^\dagger_\uparrow\hat \alpha_\uparrow
  \,\big]\,
  \rangle\\*\nonumber
 &=& -2\tilde\mathcal{N}^4{\rm Im}(\Theta)^2\;
  \langle
  \,\big[\,\hat \alpha^\dagger_\downarrow \hat \beta_\downarrow -
   \hat \beta^\dagger_\downarrow\hat \alpha_\downarrow
  \,\big]\,
  \,\big[\,\hat \alpha^\dagger_\uparrow \hat \beta_\uparrow -
   \hat \beta^\dagger_\uparrow\hat \alpha_\uparrow
  \,\big]\,
  \rangle.
\end{eqnarray}
For the last identity, we have made use of the assumption that the
total wave function of our non-degenerate ground state can be chosen to
be real-valued with respect to the quasi-momentum basis.

From the single particle band structure in figure \ref{figure3}, we may infer 
that the single particle ground state and the energetically highest excited
single particle state of a $\sigma$ boson are superpositions of
$|Q=0\rangle_{\sigma,A}$ and
$|Q=0\rangle_{\sigma,B}$. Therefore, $\delta g_2^{\downarrow,\uparrow}$
can be expressed in terms of annihilation and creation operators of these two
eigenstates, which can
shed a light on the magnitude of $\delta g_2^{\downarrow,\uparrow}$
from an energetic viewpoint. In the following, we denote
the annihilation operators
of a $\sigma$ boson in the single particle ground and highest
excited state as $\hat \psi_{\sigma,g}$ and $\hat \psi_{\sigma,e}$,
respectively. Without loss of generality, we may write:
\begin{equation}\label{eq:array}
\left\{
\begin{array}{ll}
\hat \psi_{\uparrow,g} =sin(\theta)\hat \alpha_\uparrow+cos(\theta)\hat
\beta_\uparrow\\
\hat \psi_{\uparrow,e} =cos(\theta)\hat \alpha_\uparrow-sin(\theta)\hat
\beta_\uparrow\\
\end{array}
\right.
\hspace{1em}
\left\{\begin{array}{ll}
\hat \psi_{\downarrow,g} =sin(\theta)\hat \beta_\downarrow+cos(\theta)\hat
\alpha_\downarrow\\
\hat \psi_{\downarrow,e} =cos(\theta)\hat \beta_\downarrow-sin(\theta)\hat
\alpha_\downarrow\\
\end{array}\right.
\end{equation}
The value of $\theta$ is then determined by $\varepsilon$, $J_1$, and
$J_{2,d/s}$.
Consequently, we obtain:
\begin{eqnarray}\label{delta_g2_coherence}
\bar{n}^2\,\delta g_2^{\downarrow,\uparrow} =
 -2\tilde\mathcal{N}^4{\rm Im}(\Theta)^2\;
  \langle
  \,\big[\,\hat \psi^\dagger_{\downarrow,g} \hat \psi_{\downarrow,e} -
   \hat \psi^\dagger_{\downarrow,e}\hat \psi_{\downarrow,g}
  \,\big]\,
  \,\big[\,\hat \psi^\dagger_{\uparrow,g} \hat \psi_{\uparrow,e} -
   \hat \psi^\dagger_{\uparrow,e}\hat \psi_{\uparrow,g}
  \,\big]\,
  \rangle.
\end{eqnarray}
This result clearly shows that the magnitude of
$\delta g_2^{\downarrow,\uparrow}$ is controlled
by coherences in the inter-species reduced two-body density operator with
respect to the single particle eigenstates of lowest and highest eigenenergy.

\section*{References}
\bibliography{reference}
\bibliographystyle{iopart-num}

\end{document}